\documentclass[journal=jacsat, manuscript=article]{achemso}

\usepackage[version=3]{mhchem} 
\usepackage{textcomp,  gensymb}
\usepackage{siunitx}
\usepackage[utf8]{inputenc}
\usepackage[T1]{fontenc}
\usepackage{mathptmx}
\usepackage{amsmath}
\usepackage{graphicx}
\usepackage{caption}
\usepackage{subcaption}
\usepackage{amsfonts}
\usepackage[section]{placeins} 
\usepackage{amssymb}
\usepackage{accents}
\usepackage{float}
\usepackage{graphics}
\usepackage{multicol}
\usepackage{tablefootnote}
\usepackage{threeparttable}

\makeatother

\newcommand{\response}[1]{{\color{black}{#1}}}

\author{Samhita Kattekola}
\affiliation{Department of Chemical Engineering,  The City College of New York,  CUNY,  New York,  NY 10031,  USA}
\author{Vinod M. Menon}
\altaffiliation{Department of Physics,  The City College of New York,  CUNY,  New York,  NY 10031,  USA}
\author{Alexander Couzis}
\affiliation{Department of Chemical Engineering,  The City College of New York,  CUNY,  New York,  NY 10031,  USA}
\author{Ilona Kretzschmar}
\affiliation{Department of Chemical Engineering,  The City College of New York,  CUNY,  New York,  NY 10031,  USA}
\email{kretzschmar@ccny.cuny.edu}
\phone{+1 212 6506769}
\fax{+1 212 6506660}
\affiliation{Department of Chemical Engineering,  The City College of New York,  CUNY,  New York,  NY 10031,  USA}

\title[An \textsf{achemso} demo]
  {
 Ellipsometric Identification of Transition from Layered Metal-dielectric Film to Hyperbolic Metamaterial 
 }

\abbreviations{IR, NMR, UV}

\begin{document}








\begin{abstract}

Hyperbolic Metamaterials  (HMMs) continue to be intriguing due to their applications in super resolution imaging and spontaneous emission control.
One of the successful realizations of HMMs is a layered metal-dielectric film.
Despite the extensive knowledge in thin film technology and the promises of HMM's applications,  the scale up and practical utilization of HMMs have not yet occurred. 
A general design approach is needed to predict the transition of a layered structure into an HMM. In this work, effective medium approximation and transfer matrix method are combined to determine the transition and validated by spectroscopic ellipsometry measurements on a predefined HMM structure made of silver and alumina.
Four interdependent design parameters are explored: thicknesses of metal and dielectric layers,  transition wavelength,  and minimum number of periods required for a layered metal-dielectric film to display hyperbolic dispersion.
The findings are presented as a practical engineering design chart, similar to a state diagram, that can be extended to other combinations of materials.

\end{abstract}


\section{Introduction}

Interactions between metals and dielectrics have been utilized to create interesting optical effects since ancient Roman times. \cite{RomanCup}
The history of thin inorganic films is particularly rich,  spanning nearly 5000 years of technological innovation. \cite{5000yrsofInorganicFilms, InorganicFilmsHistory} 
The advent of vacuum technology has further enriched the utilization of thin films.\cite{InorganicFilmsHistory, ThinFilmTechnologyHandbookVacuumEvaporationChapter}
Although practically used for a variety of applications ranging from decoration to mirrors throughout history,  the theoretical framework of optics for metal–dielectric systems was established by Paul Drude only in the late 19th century.\cite{drudeTheoryOptics} 
Subsequently,  the discovery and innovation of modern optical systems evolved exponentially,  leading  (but not limited) to hyperbolic metamaterials in the early 21st century.\cite{EMWavePropogationIndefinite}

Metamaterials,  initially theorized as “left-handed” materials in 1968 by Vesalago, \cite{veselagoElectrodynamicsMaterialsNegative} are non-naturally occurring materials with negative electric permittivity,  $\varepsilon$,  and negative magnetic permeability,  $\mu$. 
The past two decades have seen an exponential rise in the discovery of metamaterials.\cite{liuMetamaterials} 
The coining of the term “metamaterials” and evolution of metamaterials,  from the earliest realization in the form of split-ring resonators in the 1990s to modern metasurfaces,  has been reviewed by Ali et al.\cite{aliMetamaterialsMetasurfaces}. 
Metamaterials have been fashioned into metadevices\cite{Metadevices, haOptoelectronicMetadevices2024, pitchappaTerahertzMEMSMetadevices2021, heReviewMicroNanoProcessing2025} and are starting to be explored in the colloidal space.\cite{ColloidalMetamaterials, wangColloidalNanoparticleMetasurfaces2025, kimAdvancesColloidalBuilding2023, mannColloidalPlasmonicNanocubes2021, contiColloidalPlasmonicMetasurfaces2024, heReviewMicroNanoProcessing2025}.

Hyperbolic Metamaterials  (HMMs),  are a class of metamaterials that have garnered interest because of their applications in super-resolution imaging\cite{MetamaterialsandImaging} and spontaneous emission control\cite{noginovControllingSpontaneousEmission2010} among others.
HMMs comprise metal and dielectric domains arranged in a specific sub-wavelength configuration to tailor desired optical properties.
The theory,  design, and plasmonic character have been reviewed extensively by Poddubny et al.\cite{poddubnyHyperbolicMetamaterials},  Shekhar et al.\cite{shekharHyperbolicMetamaterials} and most recently by Ramírez-Aragón et al.\cite{ramirezStratifiedPlasmonicResonances} 

HMMs are uniaxial anisotropic materials with a tensorial effective electric permittivity as shown in \response{eq }\ref{eq:permittivitytensor},  whose parallel  ($\varepsilon_{\parallel}$) and perpendicular  ($\varepsilon_{\perp}$) components have opposite signs.\cite{EMWavePropogationIndefinite}. 
This is in contrast to isotropic materials where $\varepsilon_{\parallel}$ = $\varepsilon_{\perp}$,  or anisotropic materials where $\varepsilon_{\parallel}$ and $\varepsilon_{\perp}$ are both either positive or negative.
The dispersion of light in a medium described by the optical isofrequency curve  (IFC) is given in \response{eq }\ref{eq:isofrequencycurve},  where $k_{x}, k_{y}$,  and $k_{z}$ are the x,  y,  and z components of the wave vector,  $k$,  respectively,  $\omega$ is the angular frequency, and $c$ is the speed of light.
HMMs display hyperbolic dispersion,  i.e., the shape of the resulting IFC for an HMM is an open hyperboloid surface,  in contrast to the usually observed closed ellipsoidal surface. 
\begin{equation}
    \mathbf{\varepsilon_{eff}}=\begin{bmatrix}
        {\varepsilon_{xx}} & 0 & 0\\0 & {\varepsilon_{yy}} & 0\\0 & 0 & {\varepsilon_{zz}}
        \end{bmatrix}
        =\begin{bmatrix}
        {\varepsilon_{\parallel}} & 0 & 0\\0 & {\varepsilon_{\parallel}} & 0\\0 & 0 & {\varepsilon_{\perp}}
    \end{bmatrix}
    \label{eq:permittivitytensor}
\end{equation}
\begin{equation}
\frac{k_{x}^{2}+k_{y}^{2}}{\varepsilon_{\perp}}+\frac{k_{z}^{2}}{\varepsilon_{\parallel}}=\frac{\omega^2}{c^2}
\label{eq:isofrequencycurve}
\end{equation}

As the electric permittivity changes with wavelength,  the value of the real part of $\varepsilon_{\parallel}$ and $\varepsilon_{\perp}$ determines the regime in which a material has elliptical vs. hyperbolic dispersion. \cite{krishnamoorthyscience}
The material has hyperbolic dispersion when either $\varepsilon_{\parallel}$ or $\varepsilon_{\perp}$ is negative. 
When $\varepsilon_{\parallel}$ is positive and $\varepsilon_{\perp}$ is negative,  the IFC is a one-sheet hyperboloid  (Type I HMM),  whereas a positive $\varepsilon_{\perp}$ with negative $\varepsilon_{\parallel}$ produces a two-sheet hyperboloid IFC (Type II HMM).\cite{shekharHyperbolicMetamaterials}

\begin{figure}
    \centering
        \includegraphics[width=0.5\linewidth]{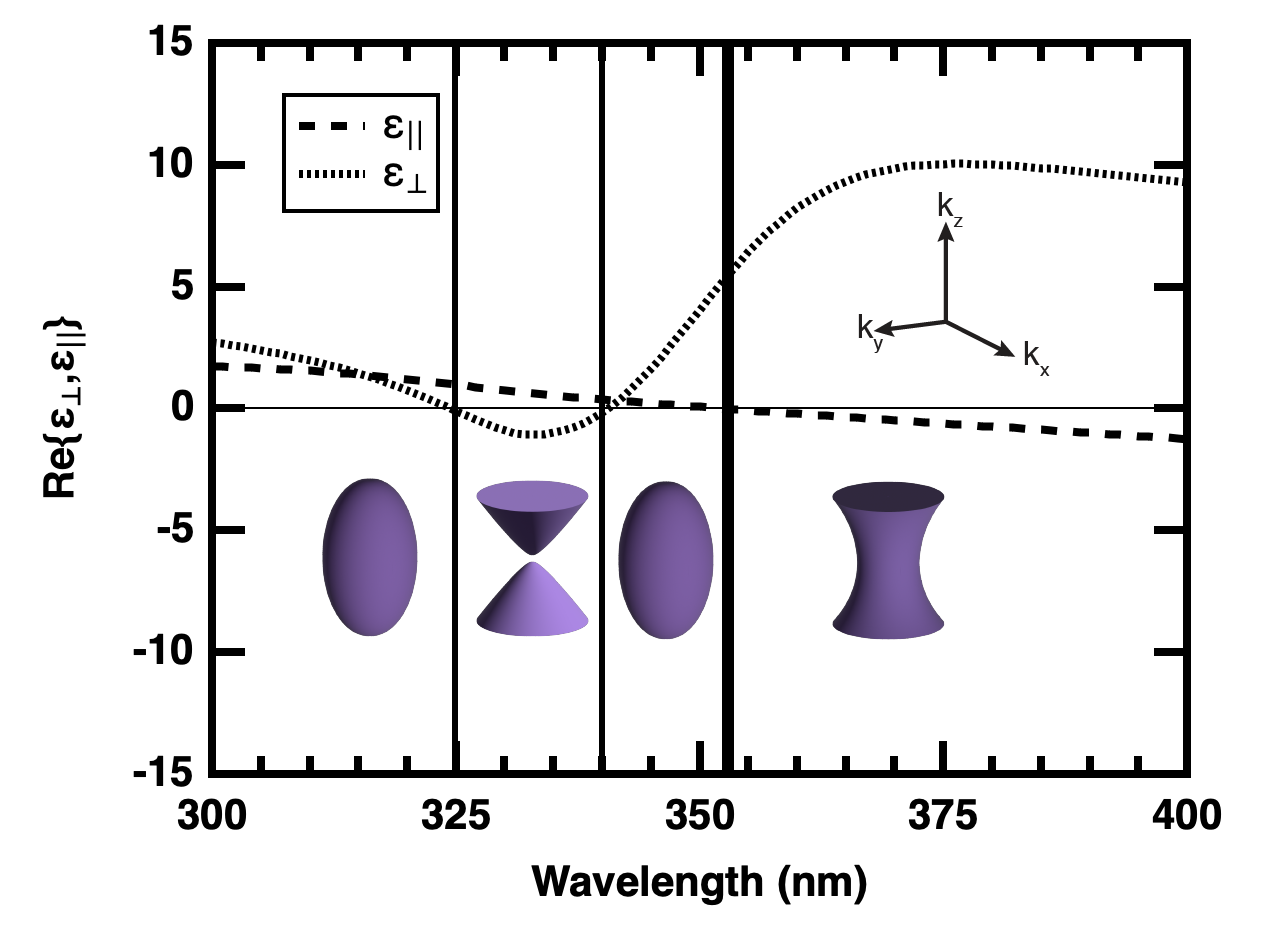} 
    \caption{Evolution of the real part of the electric permittivity components from material with elliptical dispersion to hyperbolic metamaterial generated using optical constants obtained in this work for metal thickness of 10 nm and dielectric thickness of 10 nm ($\Phi=0.5$).
    $\varepsilon_{\perp}$ and $\varepsilon_{\parallel}$ are represented by dotted and dashed lines, respectively. 
    The three vertical solid lines indicate the wavelengths at which the material first transitions from elliptical dispersion to Type I HMM,  back to elliptical dispersion and and then to Type II HMM (thicker vertical solid line, 
    $\lambda_{HMM}$),  respectively. The insets show graphical representations of the iso-frequency contour  (IFC) for the respective regimes and the wave vector reference axis orientation.} \label{fig:epsilon}
\end{figure}

\ref{fig:epsilon} shows the real part of the components of electric permittivity as a function of wavelength for a uniaxial anisotropic material accompanied by graphical representations of the IFC for each regime.
The dashed line shows the variation in $\varepsilon_{\parallel}$,  whereas the dotted line shows the variation in $\varepsilon_{\perp}$ as a function of wavelength. 
The transition points, indicated by three vertical solid lines, represent the wavelengths at which the IFC transitions first from elliptical dispersion to Type I HMM ($\lambda_{ellip-Type I}\approx325$ nm), then back to elliptical dispersion ($\lambda_{Type I-ellip}\approx340$ nm) and last from elliptical dispersion to Type II HMM  ($\lambda_{ellip-Type II}\approx353$),  respectively.
The transition wavelengths and regimes observed strongly depend on the realization of the HMM. 
It is possible to have all three transitions as shown in \ref{fig:epsilon}, two transitions of either elliptical to Type I to Type II HMM or elliptical to Type II to Type I HMM, or only one transition of either elliptical to Type I or elliptical to Type II .\cite{cortesQuantumNanophotonicsUsing2012}
One of the common realizations for HMMs is a film with alternating metal and dielectric layers, the effective electric permittivity of which depends on the fill fraction, $\Phi$.\cite{shekharHyperbolicMetamaterials} 
Metal-dielectric films which exhibit Type II hyperbolic dispersion  (hereafter referred to as HMM character) are the focus of this research. 
\ref{fig:epsilon} is such an example where $\Phi=0.5$.
The wavelength at which the transition to Type II occurs is labeled as $\lambda_{HMM}$ (thick vertical solid line in \ref{fig:epsilon}).

Abundant literature exists on growing layered thin-films that emphasizes the interest in these novel materials.\cite{ThinFilmTechnologyHandbookVacuumEvaporationChapter} 
Research over the past decade has focused on developing novel Type II HMMs,  characterizing their optical response,  identifying potential applications, and generally pushing the limits of HMMs' usage.
The first experimental demonstration of a Type II HMM was reported by Liu et al.\cite{firsthyperlens} in the form of a hyperlens that surpassed the diffraction limit of light using a curved surface made of silver and alumina layers.
Work by Hoffman et al.\cite{hoffmanNegativeRefraction} showed negative refraction by layering doped semiconductor films.
Krishnamoorthy et al.\cite{krishnamoorthyscience} introduced topological transitions using silver and titanium oxide layers showing an increase in the spontaneous emission rate of quantum emitters in such structures.
For additional combinations,  designs,  and applications the reader is referred to the reviews by Poddubny et al.\cite{poddubnyHyperbolicMetamaterials} and Cortes et al.\cite{cortesQuantumNanophotonicsUsing2012}.

The common design parameters in the literature for the fabrication of HMMs with alternating metal-dielectric layers are: metal thickness,  $h_m$,  dielectric thickness,  $h_d$,  the electric permittivity of the metal and dielectric material,  $\varepsilon_m$ and $\varepsilon_d$,  respectively,  the fill fraction of metal in the system,  $\Phi$ \response{(eq \ref{eq:fillfraction})}, and the transition wavelength for the system to become an HMM,  $\lambda_{HMM}$.
The design parameters are generally chosen based on effective medium theory  (EMT) calculations. \cite{MaxwellGarnet} Effective medium implies that the film made of metal-dielectric layers behaves as a homogeneous material with an effective electric permittivity; $\varepsilon_{\parallel}$ and $\varepsilon_{\perp}$ are then calculated using \response{eq \ref{eq:epara} and eq \ref{eq:eperp}, respectively.}

\response{\begin{equation}
    \Phi=\frac{h_m}{h_m+h_d}
    \label{eq:fillfraction}
\end{equation}
\begin{equation}
\varepsilon_{\parallel}=\Phi\varepsilon_m+ (1-\Phi)\varepsilon_d
\label{eq:epara}
\end{equation}
\begin{equation}
\varepsilon_{\perp}=\frac{\varepsilon_m\varepsilon_d}{\Phi\varepsilon_d+ (1-\Phi)\varepsilon_m}
\label{eq:eperp}
\end{equation}}

Essentially,  for a given combination of metal and dielectric,  $\lambda_{HMM}$ is a function of $\Phi$  (\ref{fig:epsilon}).
However,  $\lambda_{HMM}$ and $\Phi$ are insufficient in identifying the number of periods at which the material transitions from a metal-dielectric layered structure into an HMM. 
For instance, consider \ref{fig:epsilon}, which has a $\Phi=0.5$, predicts the same solution ($\lambda_{HMM}\approx353$ nm) for a 5:5 and a 10:10 nm metal:dielectric layer ratio. 
However, the 5:5 structure will likely need more periods to become an HMM.
EMT simply does not take into account the impact of the number of periods in the film or total film thickness,  both of which impact whether the resulting film is an HMM.\cite{elserNonlocalEffectsEffectivemedium2007, cortesQuantumNanophotonicsUsing2012}
This represents a limitation of EMT's applicability in  the predictive design of HMMs and increases the reliance on complex optical setups to identify if the material produced exhibits HMM character only after it is fabricated.\cite{poddubnyHyperbolicMetamaterials}

A typical work flow found in the literature involves choosing the size and combination of materials  ($h_m$ and $h_d$) most suited for the desired application. 
Consequently,  $\lambda_{HMM}$ is obtained by the utilization of optical constants available in the literature,  EMT calculations  (\response{eq  \ref{eq:fillfraction} - 5}), and construction of the corresponding \ref{fig:epsilon}.\cite{krishnamoorthyscience} 
Then, the constituent metal and dielectric films are grown, characterized to obtain actual thicknesses and optical constants using ellipsometry, and $\lambda_{HMM}$ is recalculated for the actual materials made. 
Next,  samples of at least four periods  (but also as many as eight or 200 periods have been reported)\cite{firsthyperlens,  Kelly:16,elserNonlocalEffectsEffectivemedium2007,sukhamInvestigationEffectiveMedia2019a} are grown and it is not apparent from the literature how the number of periods is chosen.
Subsequently,  optically active materials such as quantum dots are coupled to the metamaterial to investigate its optical properties.\cite{galfskyActiveHyperbolicMetamaterials2015, shekharHyperbolicMetamaterials, luActiveHyperbolicMetamaterials2018}
It is only in the last step,  which requires complicated optical and photonics setups,  that the material is confirmed to be an HMM,  posing practical limitations on the utilization of HMM.
The review by Lee et al.\cite{leeHyperbolicMetamaterialsFusing2022} summarizes complex experiments and calculations that have been performed for HMM realizations and concludes with suggestions to extend the range of practical applications of HMMs.

The complexity in HMM characterization methods mentioned above has led to a search for simpler approaches such as reflectometry and ellipsometry to identify HMM character.\cite{tumkurPermittivityEvaluationMultilayered2015a, mroczynskimagnetron, krishnamoorthyNonVolatileChalcogenideSwitchable2018, diltsLowMSEExtractionPermittivity2019, oatesCharacterizationPlasmonicEffects2011, langGoldsiliconMetamaterialHyperbolic2013}
Ellipsometry, \cite{tompkinsMetal, 
hilfikerabsorbing} one of the most commonly used techniques for characterization of inorganic and organic thin films,  has only sparingly been used for HMM characterization.
The complexity in modeling HMM using ellipsometry primarily arises from difficulty in modeling the out-of-plane electric permittivity component. 
Several works have focused on estimating the out-of-plane permittivity and fitting them to the experimental data as uniaxial models.\cite{tumkurPermittivityEvaluationMultilayered2015a, mroczynskimagnetron, krishnamoorthyNonVolatileChalcogenideSwitchable2018, diltsLowMSEExtractionPermittivity2019, Kelly:16, zhangRobustExtractionHyperbolic2018} 
Prominent work that experimentally determines both in- and out-of-plane permittivity is that by Tumkur et al.\cite{tumkurPermittivityEvaluationMultilayered2015a},  Kelly et al.\cite{Kelly:16},  and Zhang et al.\cite{zhangRobustExtractionHyperbolic2018}.
They assume that the materials being characterized are uniaxial homogeneous materials and compare them to theoretical calculations to discern the effective permittivity of the HMM.
Tumkur et al.\cite{tumkurPermittivityEvaluationMultilayered2015a} and Kelly et al.\cite{Kelly:16} utilize anisotropic and general spectroscopic ellipsometry measurements,  respectively,  whereas Zhang et al.\cite{zhangRobustExtractionHyperbolic2018} employs isotropic spectroscopic ellipsometry measurements to obtain the effective permittivity.
A key distinction in Zhang et al.\cite{zhangRobustExtractionHyperbolic2018} is the measurement of the sample itself as they use total internal reflection ellipsometry,  requiring an index-matching prism\cite{huOttoconfig} to increase sensitivity of the out-of-plane interactions to robustly extract the out-of-plane permittivity coefficients.

Here,  we posit the question: Can ellipsometry,  a fundamentally far-field method,  be used to distinguish an HMM from a multilayer film?
If so,  can it be used to predict the transition of a layered metal-dielectric film into a hyperbolic metamaterial? 
To answer these questions,  a previously verified and known hyperbolic metamaterial\cite{galfskyActiveHyperbolicMetamaterials2015} comprising silver and alumina layers with seven periods  (\textbf{7P}) is fabricated. Additionally,  the one period  (\textbf{1P}) and four period analogues  (\textbf{4P}) are also grown. All films are characterized and analyzed using isotropic spectroscopic ellipsometry measurements. AFM is used to independently confirm film thicknesses.
Subsequently,  ellipsometry data is compared to numerical simulation predictions based on a \text{2x2} transfer matrix method  (TMM). Next,  the simulations are used to predict the ellipsometric response for the \ce{Ag-Ge}/\ce{Al2O3} system as a function of number of periods revealing the threshold for HMM behavior.
A state diagram is constructed that allows prediction of the HMM transition using four interdependent design parameters: metal layer thickness,  $h_{m}$,  dielectric layer thickness,  $h_{d}$,  minimum number of periods of metal-dielectric layers in the film,  $P_{HMM,min}$,  and the wavelength at which the transition into HMM occurs,  $\lambda_{HMM}$. 
The resulting design chart enables tailoring of hyperbolic metamaterials for specific applications such as hyperlens and enhanced spontaneous emission.

\section{Experimental and Simulation Details}
\subsection{Material Fabrication}

Samples are fabricated using an e-beam evaporator  (AJA International Orion 8E Evaporator) maintained at $P < 10^{-8}$ torr and equipped with a quartz crystal microbalance to monitor thickness (reported as nominal thickness).
All films are grown at an average rate of $0.3 \textup{~\AA}/s$. 
Alumina  (\ce{Al2O3}),  silver  (\ce{Ag}),  and germanium  (\ce{Ge}) targets used are purchased from Kurt J Lesker. 
Single-side polished  (SSP) silicon wafers purchased from University Wafer,  Inc. are used as substrates. 
Substrates are cleaned by submerging them in a mixture of Nochromix\textregistered\,  and sulfuric acid for two hours,  followed by copious rinsing with deionized water and drying in an oven at 70$\degree$C. 
Any sample containing silver requires deposition of a germanium seed layer  (nominal thickness of 2 nm) to allow layered film growth.\cite{UltrasmoothSilverFilms}
Additionally,  whenever the last layer of the sample is silver,  a protection layer of alumina  (nominal thickness of 6 nm) is grown to prevent silver oxidation after removal from the evaporator.
\begin{figure}
    \centering
        \includegraphics[width=\linewidth]{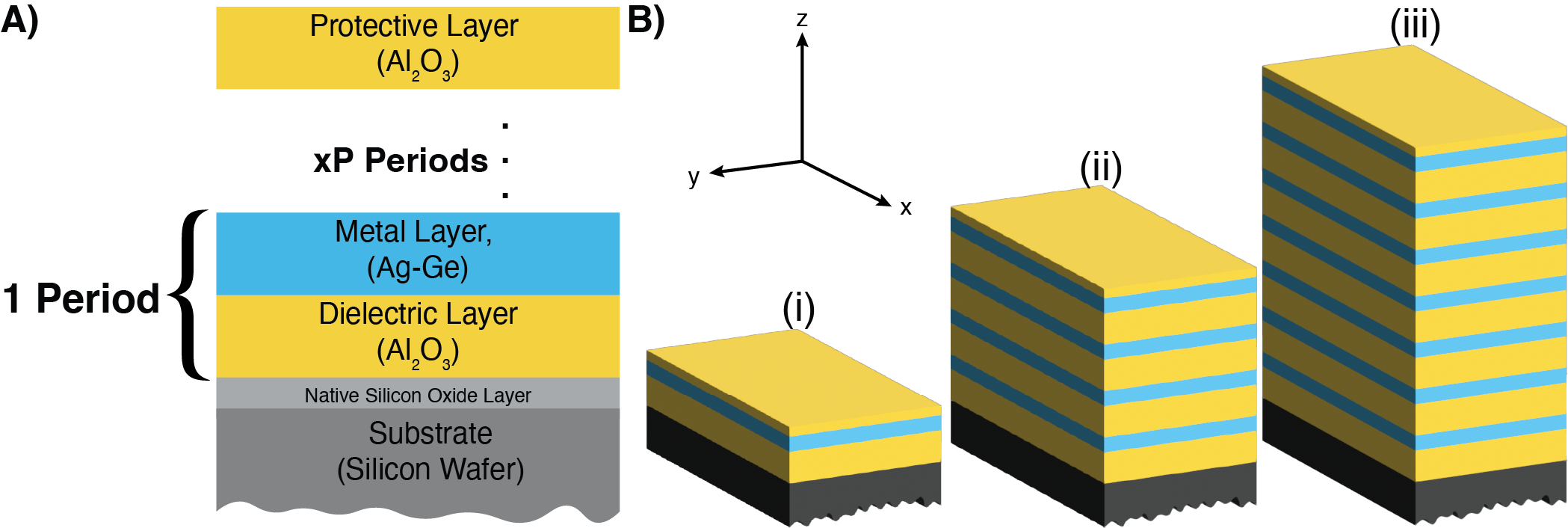} 
    \caption{Schematics of layered films.
 (A) Cross-section of fabricated periodic layered metal-dielectric film. One period consists of a dielectric layer  (DL,  yellow) of  alumina  (\ce{Al2O3}) and a metal layer  (ML,  blue) of  silver grown on a seed layer of germanium  (\ce{Ag-Ge}). All films are capped with a protection layer  (PL,  yellow) of \ce{Al2O3}. A silicon wafer with a native oxide layer is used as substrate  (gray and light gray,  respectively).
 (B) Schematic of (i) \textbf{1P}-,  (ii) \textbf{4P}-,  and (iii) \textbf{7P} films. 
} 
\label{fig:SamplesControlsXsectionHMM}
\end{figure}

\begin{table}[]
 \caption{Sample nominal thicknesses read from e-beam evaporator thickness monitor in nm. $N$ is the number of \ce{Ag-Ge}/\ce{Al2O3} periods.}
  \label{tab:samples}
   \begin{threeparttable}
 \centering
   \begin{tabular}{|c|c|c|c|c|c|}
   \hline
         Sample	&$N$	&Total Alumina$^a$ &	Total Germanium	&Total Silver	&Total$^a$\\
         \hline
\textbf{1P}&1&21&2&10&33\\\textbf{4P}&4&66&8&40&114\\\textbf{7P}&7&111&14&70&195\\
         \hline
    \end{tabular}
        \begin{tablenotes}
       \item [$a$] Value includes 6 nm \ce{Al2O3} protection layer. 
     \end{tablenotes} 
 \end{threeparttable}
\end{table}

\ref{fig:SamplesControlsXsectionHMM} shows schematics of the layerd films fabricated on silicon wafer substrates. 
Three samples of each film are grown.
The cross-section of the film is shown in \ref{fig:SamplesControlsXsectionHMM}A. One period of the film has a nominal thickness of $27$ nm,  comprising $15$ nm alumina,  $2$ nm seed layer of germanium and $10$ nm of silver.  
Since the period ends in silver, a 6 nm protection layer of alumina is deposited on all layered films.
\ref{fig:SamplesControlsXsectionHMM}B shows the three layered metal-dielectric films comprising (i) one (\textbf{1P}),   (ii) four (\textbf{4P}), and (iii) seven (\textbf{7P}) periods. 
\ref{tab:samples} summarizes the nominal thicknesses for all samples. 
In addition, alumina films with five nominal thicknesses (3,  6,  10,  15,  and 20 nm) and germanium-silver-alumina films with four silver nominal thicknesses (5,  10,  15,  and 20 nm) are grown for ellipsometry calibration.

\subsection{Ellipsometry Measurements}

A J.A. Woollam Co, Inc. V-VASE\textregistered\ Ellipsometer equipped with WVase\textregistered\, Software\cite{WVASE} is used to conduct variable angle spectroscopic ellipsometry  (SE). 
Isotropic ellipsometry measurements for all samples are done over a wavelength range of $300-2000$ nm with a step size of $30$ nm  ($10$ nm step for silicon wafer substrate) for two incident angles  (65$\degree$ and 75$\degree$). 
The ellipsometer measures the ratio of Fresnel reflection coefficients,  $\mathfrak{R}_p$ and $\mathfrak{R}_s$,  for p- and s-polarization,  respectively.
The measurement is reported either as $\Psi$ and $\Delta$ or $\rho$,  which are interrelated as given in \response{eq }\ref{eqn:ellipsomtry eqn}.
\begin{equation}
  \rho=\frac{\mathfrak{R}_p}{\mathfrak{R}_s} = \tan{\Psi}e^{i\Delta}  
  \label{eqn:ellipsomtry eqn} 
\end{equation}  

The J.A. Woollam Co, Inc. CompleteEASE\textregistered\, Software\cite{CompleteEASE} is used for modeling of optical constants and for regression analysis to obtain layer thicknesses and a goodness-of-fit, which is reported as root mean square error (MSE).\cite{CompleteEASE}
The description of MSE and its relation to $\Psi$ and $\Delta$ are shown in SI Section I.2.1.

The modeling of the films is done in four consecutive stages.
The cleaned silicon wafer substrate is modeled first using the native oxide model provided by the Woollam CompleteEASE\textregistered\, Software  (MSE and native oxide thickness reported in SI Section I.2.2).
This model is used as a substrate layer without adjustment for all further modeling.
Next, the alumina layer is modeled by modifying the CompleteEASE\textregistered\,  built-in Cauchy model for \ce{Al2O3} to reflect the optical properties of alumina deposited using e-beam evaporation.\cite{aluminafilmRI, houskaOverviewOpticalProperties2012}
The silver layer is defined as a composite \ce{Ag-Ge} layer and modeled using the CompleteEASE\textregistered\,  general oscillator model. 
A combination of Drude,  Lorentz and Cody-Lorentz oscillators are used to reflect the layer of silver with the germanium seed layer.\cite{SilveronGeRI}.
The reported optical constants for individual layers of \ce{Al2O3} (Fig. S4) and \ce{Ag-Ge} (Fig. S6) are checked to be consistent with those reported in the literature. 
MSE and thickness calibration details for the individual layers are provided in SI Section I.2.3 and I.2.4, respectively. 
Based on the optical constants and layer thicknesses, the parallel and perpendicular components of the electric permittivity tensor for \ce{Al2O3} and \ce{Ag-Ge} system studied here, are predicted using Eqn. 3 (SI Section I.3, Fig. S7).
Subsequently, the layered films are modeled as a multilayer model with a silicon wafer substrate,  and alternating \ce{Al2O3} and \ce{Ag-Ge} layers culminating in the \ce{Al2O3} protection layer (see \ref{fig:SamplesControlsXsectionHMM}). 
Here, a single period comprises two layers, \ce{Al2O3} and \ce{Ag-Ge}.
For films with more than one period,  a technique known as parameter coupling is employed to keep the number of fitting parameters in the model constant.\cite{ParameterCoupling}
Parameter coupling is a built-in feature of CompleteEASE\textregistered\,,  wherein the relationship between two layer parameters is defined. 
It assumes that every alternative repeating layer has the same thickness and optical constants.
This assumption,  truly represents the periodic multilayer character of the film as fabricated.
The reduction in fitting parameters simplifies the system and also reduces error propagation from multiple fit parameters.
The data for three independently fabricated samples of each film are simultaneously fitted for both angles of incidence and reported in \ref{tbl:thick}.

\subsection{AFM Measurements}

Atomic Force Microscopy  (AFM) is used on selected calibration and film samples to confirm thickness measurements predicted by ellipsometry.
A Bruker Dimension FastScan Atomic Force Microscope is used in contact mode with a ContAI-G AFM Probe from BudgetSensors. 
To enhance accuracy of the AFM thickness measurement, a subset of films is deposited on a $4$ $\mu$m silica particle submonolayer that is subsequently lifted off. 
The silica particles shadow the substrate from the depositing material and leave behind an empty area on the substrate. 
The boundary between the shadowed area and the deposited film allows accurate measuring of the film thickness  (see SI Section I.1.2).
AFM measurements are reported along with ellipsometry calibrations data in SI Section I as well as with \textbf{1P}, \textbf{4P}, and \textbf{7P} film data in \ref{tbl:thick}.

\subsection{Numerical Simulations}

MATLAB\textregistered\, \cite{MATLAB2025aPCT} is used to simulate the real (Re$\{\rho\}$) and imaginary (Im$\{\rho\}$) parts of the ellipsometric parameter,  $\rho$,  for combinations of \ce{Ag-Ge} and \ce{Al2O3} layer thicknesses ranging from 1 to 20 nm and 1 to 40 nm, respectively,  with a resolution of 0.01 nm. 
The wavelength range for the simulation is 300 to 2000 nm with a step size of 1 nm.
In order to discern HMM character, Re$\{\rho\}$ is calculated for all combinations at $\lambda_{HMM}$ for films with 1 to 150 periods.
Re$\{\rho\}$ values from films with 100 to 150 periods are averaged and reported as Re$\{\rho _\infty\}$.
The average difference between Re$\{\rho _\infty\}$ and the highest Re$\{\rho\}$ value for the entire parameter space investigated is reported as the confidence interval $\delta$Re$\{\rho\}= \pm0.035$.
The Transfer Matrix Method  (TMM) with \text{2x2} Abeles' formulation as outlined by Born and Wolf\cite{bornPrinciplesOpticsElectromagnetic1999} is utilized. 
The optical constants for \ce{Ag-Ge} and \ce{Al2O3} obtained from ellipsometry measurements are used in the simulations.
Details of the calculations are given in SI Section II.

\section{Results and Discussion}

Three types of films with one  (\textbf{1P}),  four  (\textbf{4P}) and seven  (\textbf{7P}) periods on silicon wafer substrates are investigated. 
A multilayer model is constructed to characterize and determine the thickness of individual layers constituting the \textbf{1P} film.
The accuracy of the multilayer model is validated with AFM measurements.
Then,  the model is extended to the \textbf{4P} and \textbf{7P} films using parameter coupling to assess the model's ability to correctly represent the films with increased number of periods. 
Numerical simulations are performed to validate the experimental findings and to discern the emergence of the HMM character.  
A universal HMM design chart is build for the \ce{Al2O3}/\ce{Ag-Ge} system that reveals the minimum number of periods, $P_{HMM,min}$, and transition wavelength, $\lambda_{HMM}$, needed for an \ce{Al2O3}/\ce{Ag-Ge} system to exhibit HMM character.

\subsection{Ellipsometry Characterization of Layered Films}

\textbf{1P},  \textbf{4P},  and \textbf{7P} samples grown on silicon wafers (\ref{fig:SamplesControlsXsectionHMM}) are characterized using spectroscopic ellipsometry (SE) employing multilayer modeling and parameter coupling together with the optical constants obtained from modeling \ce{Al2O3} and \ce{Ag-Ge} films (described in experimental details and in SI Section I).

\subsubsection{\textbf{1P} Film Modeling}

\begin{figure}[h!]
    \centering
        \includegraphics[width=0.5\linewidth]{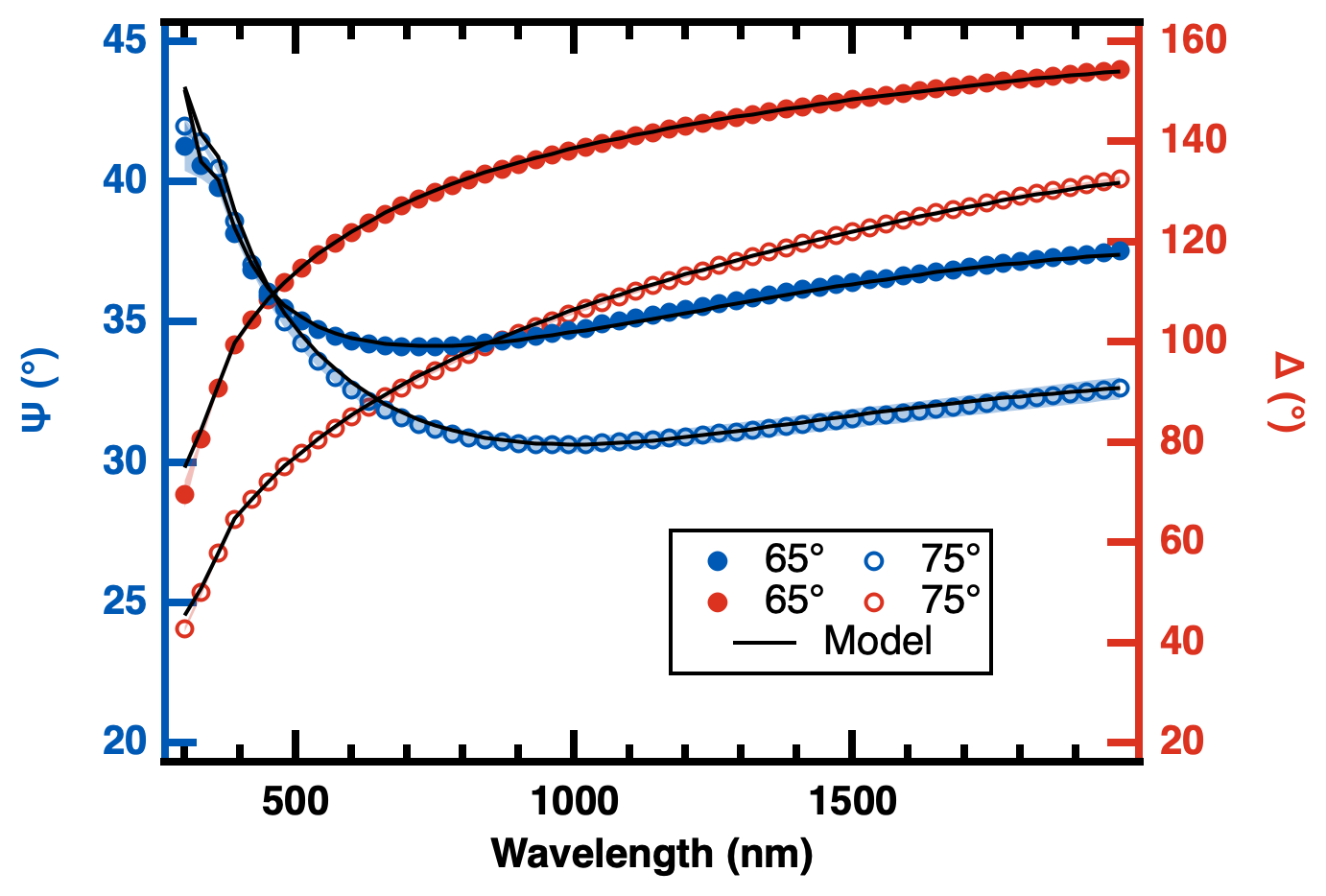} 
    \caption{Ellipsometric parameters $\Psi$  (blue,  left axis) and $\Delta$ (red,  right axis) as a function of wavelength for the \textbf{1P} film obtained at 65$\degree$  (closed circles) and 75$\degree$  (open circles) incident angles. 
    One standard deviation obtained from averaging three independently grown samples is represented as shaded area but is smaller than the marker size throughout. 
    Solid lines represent the best multilayer model fit obtained using CompleteEASE\textregistered\, Software  (see text). 
     }
    \label{fig:1P}   
    \hfill
\end{figure}

\ref{fig:1P} shows the ellipsometric parameters,  $\Psi$ and $\Delta$,  as blue and red circle markers,  respectively,  and their corresponding multilayer model fits  (solid lines) for the \textbf{1P} film on a silicon wafer as a function of wavelength,  $\lambda$.
The shaded blue and red regions  (smaller than marker size) represent one standard deviation for the average of three samples.
The filled and open markers correspond to incident angles of 65$\degree$ and 75$\degree$,  respectively. 
The multilayer model is comprised of three layers,  i.e.,  an alumina layer  (DL),  a silver layer  (ML),  and an additional alumina protective layer  (PL) as shown in \ref{fig:SamplesControlsXsectionHMM}Bi.
The three nominal thicknesses  (given in \ref{tab:samples}) are used as initial guesses for iterative modeling. 

The first row of \ref{tbl:thick} shows the thickness values obtained from the \textbf{1P} film fitting. 
The low overall MSE of 10.0 validates the use and accuracy of the multilayer model for the \textbf{1P} film.
The total thickness of $40.8 \pm 0.3\, $nm predicted by the multilayer model of the \textbf{1P} film  (Column 6 in \ref{tbl:thick}) agrees with the AFM measured thickness of $39.0 \pm 0.8\, $nm  (Column 7 in \ref{tbl:thick}) and independently validates the \textbf{1P} model and optical constants allowing calculation of the \ref{fig:epsilon} equivalent for the system studied here (SI Section I.3, Figure S7).
Next,  the validated \textbf{1P} multilayer model is used to fit the ellipsometry data obtained from the \textbf{4P} and \textbf{7P} films.

\begin{table}[h!]
  \caption{Fitted layer thicknesses,  $h_x$,  from ellipsometry for dielectric layer  ($x=DL,\ce{Al2O3}$),  metallic layer  ($x=ML,\ce{Ag-Ge}$),  protection layer  ($x=ML,\ce{Al2O3}$),  and total film thickness  ($x = total, SE$) and Measured total thickness from AFM,  $h_{total, AFM}$. All thickness values in nm. MSE is the root mean square error reported by CompleteEASE\textregistered. 
  }
  \label{tbl:thick}
  \begin{threeparttable}
    \centering
    \begin{tabular}{|c|c|c|c|c|c|c|}
    \hline
Sample&MSE&$h_{DL,\ce{Al2O3}}$&$h_{ML,\ce{Ag-Ge}}$&$h_{PL,\ce{Al2O3}}$&$h_{total, SE}$&$h_{total, AFM}$\\
\hline
\textbf{1P}&$10.0$&$21.1\pm0.3$&$10.84\pm0.03$&$8.82\pm0.09$&$40.8\pm0.3$&$39.0\pm0.8$\\
\textbf{4P}&$15.8$&$23.7\pm0.1$&$11.17\pm0.05$&$8.2\pm0.1$&$147.6\pm0.2$&$128\pm1$\\
\textbf{7P}$^a$
&$23.2$&$25.6\pm0.2$&$14.0\pm0.1$&$9.8\pm0.2$&$284.0\pm0.7$&$209\pm1$\\
\hline
    \end{tabular}
      \begin{tablenotes}
       \item [$a$] Note that the thickness values obtained from the \textbf{1P} model  (row 1) are used as the initial guess for the \textbf{7P} data fit (see text).
     \end{tablenotes} 
 \end{threeparttable}
\end{table}

\subsubsection{\textbf{4P} and \textbf{7P} Film Modeling}

\begin{figure}[h!]
    \centering
        \includegraphics[width=1.0\linewidth]{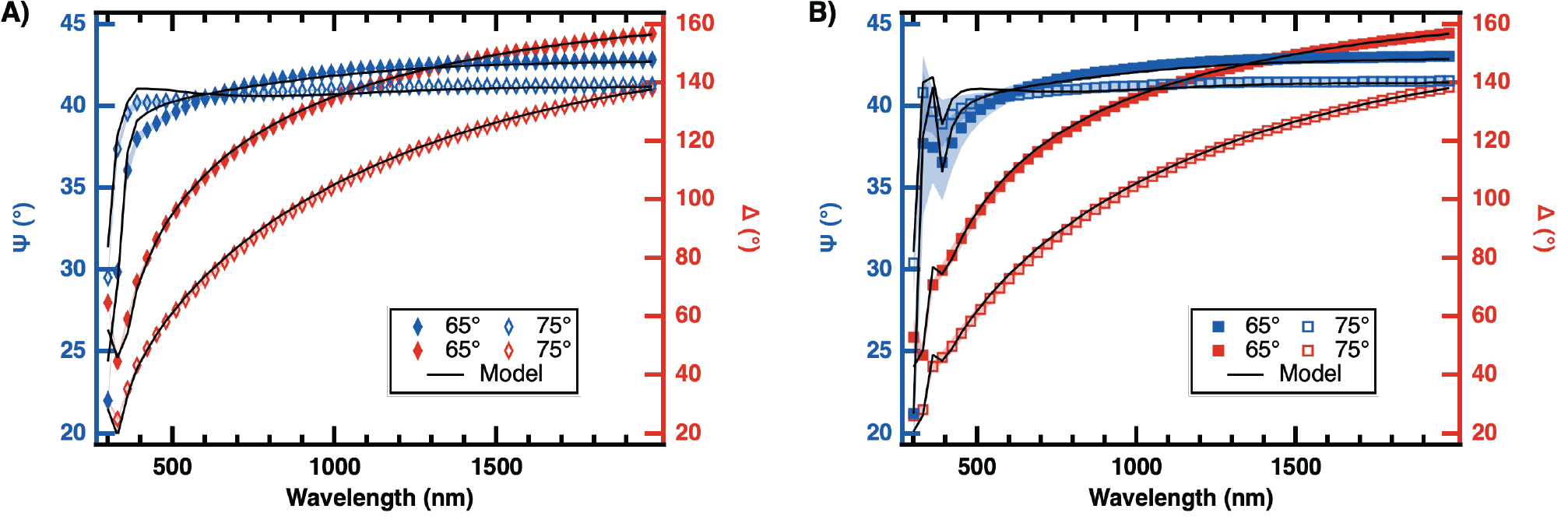} 
    \caption{Ellipsometric parameters $\Psi$  (blue,  left axis) and $\Delta$ (red,  right axis) for  (A) \textbf{4P}  (diamond markers) and  (B) \textbf{7P}  (square markers) film as a function of wavelength obtained at 65$\degree$  (filled markers) and 75$\degree$  (open markers) incident angles. 
    One standard deviation obtained from averaging three independently grown samples is represented as shaded area and is in some instances smaller than the marker size. 
    Solid lines represent the best multi-layer model fit with parameter coupling obtained using CompleteEASE\textregistered\, Software  (see text).  
     }
    \label{fig:4P-7P}   
    \hfill
\end{figure}

\ref{fig:4P-7P} shows the ellipsometric parameters,  $\Psi$ and $\Delta$,  their corresponding standard deviation,  and multilayer model with parameter coupling fits for the \textbf{4P}  (\ref{fig:4P-7P}A) and \textbf{7P}  (\ref{fig:4P-7P}B) samples using the same color and line assignment as used in \ref{fig:1P}. 
Diamond markers are used for \textbf{4P} and square markers are used for \textbf{7P}.

Comparison of \ref{fig:1P} and \ref{fig:4P-7P}A reveals that the shape of the $\Psi$ and $\Delta$ curves change substantially from the \textbf{1P} to the \textbf{4P} film.
Specifically,  the $\Psi$ curve  (blue markers) appears flipped, that is,  $\Psi$ decreases with $\lambda$ in the \textbf{1P} film data,  whereas it increases with $\lambda$ in the \textbf{4P} film data. 
Furthermore,  there is an apparent shift of the $\Delta$ curve to the right as well as the range for $\Delta$ is slightly increased in the \textbf{4P} film data starting at 20 rather than at 40.
In contrast,  the $\Psi$ and $\Delta$ curves for the \textbf{4P} and \textbf{7P} films  (\ref{fig:4P-7P}A and B) appear almost identical,  with the exception of a peak at $\lambda\approx 340$ nm in the \textbf{7P} film data. 
The apparent changes in both $\Psi$ and $\Delta$ between the \textbf{1P} and \textbf{4P} films indicate that the interaction of light with the films changes as periods are added to the film,  while the similarity of the \textbf{4P} and \textbf{7P} film responses signifies similar optical properties possibly hinting at the emergence of the HMM metamaterial character in as early as the \textbf{4P} sample as the \textbf{7P} sample has been shown to an be HMM.\cite{galfskyActiveHyperbolicMetamaterials2015} 
The \textbf{1P}-\textbf{4P} and \textbf{4P}-\textbf{7P} film data differences in $\Psi$ and $\Delta$ are shown in SI Section III.

Applying the multilayer model to \textbf{4P} and \textbf{7P} films requires an addition of either three or six \ce{Al2O3}/\ce{Ag-Ge} periods,  respectively,  to the \textbf{1P} model. 
The new layers are coupled to each other using the parameter coupling technique  (see experimental details) that assigns fitting parameters to each material and fits every layer of the materials as it occurs in the period with the same thickness parameter.  
Nominal thicknesses  (\ref{tab:samples}) are used as initial parameters for the fitting of the \textbf{4P} and \textbf{7P} film data. 
The thickness values predicted by the \textbf{4P} and \textbf{7P} parameter coupling models are shown in the second and third row of \ref{tbl:thick},  respectively.
Note,  while the \textbf{4P} fit converges with nominal thicknesses as initial parameters,  the \textbf{7P} fit provides unreasonable fitting parameters  (i.e.,  $h_{PL} = 5.9 \pm 0.2$,  $h_{ML} = 7.5 \pm 0.1$,  and $h_{DL} = 13.9 \pm 0.2$ nm,  MSE = 29.5) resulting in a total thickness of $h_{total} = 155.7 \pm 0.5$ nm,  which is much smaller  (25\%) than the \textbf{7P} film thickness measured with AFM  ($h_{AFM} = 209\pm1)$ nm.  
A second round of fitting of the \textbf{4P} and \textbf{7P} film data using the \textbf{1P} fitting parameters from \ref{tbl:thick} as initial parameters results in the same fitting parameters for \textbf{4P},  but provides a fit with optimized parameters  (third row of \ref{tbl:thick}) that shows consistent trends with the \textbf{1P} and \textbf{4P} films and also with the AFM measurement. 

Analysis of \ref{tbl:thick} reveals that both $h_{ML}$ and $h_{DL}$ are overpredicted by the model for the \textbf{4P} and \textbf{7P} films,  while $h_{PL}$ is modeled as thinner in the \textbf{4P} film and thicker in the \textbf{7P} film compared to the \textbf{1P} film.
Unsurprisingly,  comparison of total film thicknesses for the \textbf{4P} and \textbf{7P} films from ellipsometry fitting  (Column 6 in  \ref{tbl:thick}) with the actual thicknesses measured using AFM  (Column 7 in  \ref{tbl:thick}) shows also a $15\%$ and $36\%$ overprediction for the \textbf{4P} and \textbf{7P} films,  respectively.
Interestingly,  the MSE values of 15.8 and 23.2,  respectively,  for the multilayer model with parameter coupling obtained for the \textbf{4P} and \textbf{7P} films are only slightly higher than the MSE for the \textbf{1P} films (10.0),  indicating that the model still accurately represents the system.

Generally speaking,  in an isotropic material an increase in material thickness coincides with an increase in the reflectivity of the material. 
Analogously,  a more reflective film will be found to be thicker when fitted.
The fact that both $h_{ML}$ and $h_{DL}$ are overpredicted for the \textbf{4P} and \textbf{7P} films  (\ref{tbl:thick}) indicates that the films become more reflective as more periods are added than the total number of periods alone warrants. 
Transmission data  (not shown) for all three films grown on glass substrates shows the expected decrease in transmission as the number of periods in the films increases.
A key property of Type II HMMs is an increase in the overall reflectivity properties of the material due to their ability to reflect propagating waves and transmit evanescent waves. \cite{cortesQuantumNanophotonicsUsing2012,  TwoDimensionalImagingFarField2007}
Thus, the observed overprediction could be a sign of the emergence of a Type II HMM character in the \textbf{4P} film.
In order to shed light on this behavior,  numerical simulations are performed to track the change in $\rho$  (\ref{eqn:ellipsomtry eqn}) as a function of the number of periods.

\subsection{Numerical Simulations with Transfer Matrix Method}

A numerical simulation using the Abeles' transfer matrix method  (TMM) is employed to predict the ellipsometric parameter,  $\rho$,  for the \textbf{1P},  \textbf{4P} and \textbf{7P} films at an incident angle of 65$\degree$ using their respective fitting parameters from \ref{tbl:thick}. 
Re$\{\rho\}$ and Im$\{\rho\}$ from the numerical simulations show good agreement with the experimentally data  (see SI Section IV) and allow prediction of $\rho$ for films as a function of periods. 

\begin{figure}[h!]
    \centering
        \includegraphics[width=0.5\linewidth]{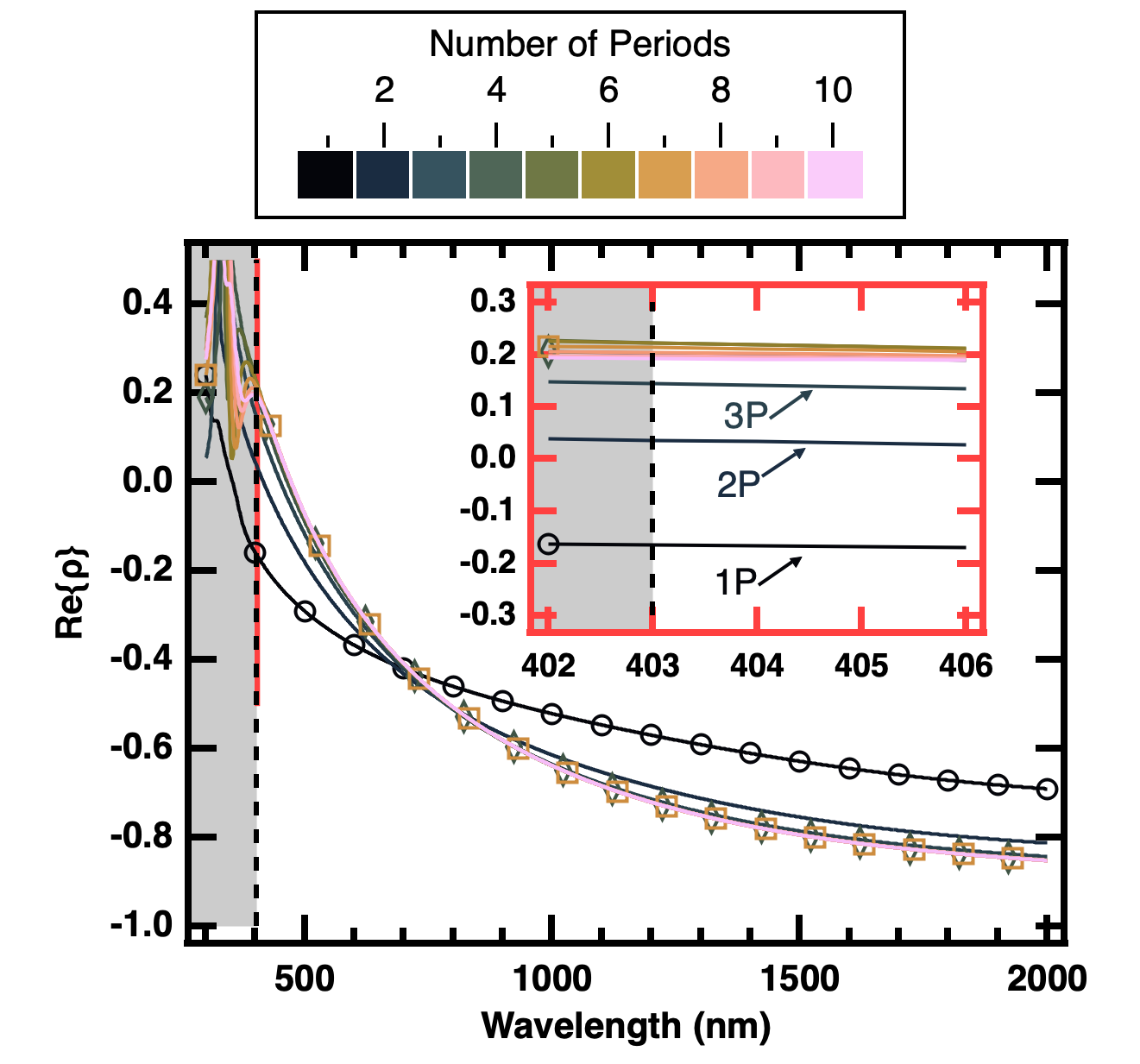} 
    \caption{Real part of $\rho$,  Re$\{\rho\}$,  as a function of wavelength calculated using the transfer matrix method. 
    Color indicates number of periods in the film.
    The dashed line indicates $\lambda_{HMM}\approx403$ nm above which the material becomes a Type II HMM. 
    Inset: Zoomed in region at $\lambda_{HMM}$ showing the convergence of Re$\{\rho\}$ starting with the \textbf{4P} film.
    Open circles, diamonds, and square markers are added to the \textbf{1P},  \textbf{4P},  and \textbf{7P} film  Re$\{\rho\}$ predictions, respectively. 
    } 
\label{fig:RhoSimGraphs}
\end{figure}

\ref{fig:RhoSimGraphs} shows the numerically simulated Re$\{\rho\}$ for a system with $h_{PL,\ce{Al2O3}} = 8.82$,  $h_{ML,\ce{Ag-Ge}} = 10.84$,  and $h_{DL,\ce{Al2O3}} = 21.1$ nm  (the \textbf{1P} fitting parameters from \ref{tbl:thick}),  which has a $\lambda_{HMM} \approx 403$ nm using the experimentally obtained optical constants (SI Section I.3, Figure S7 ).
The inset of \ref{fig:RhoSimGraphs} shows the zoomed-in part of Re$\{\rho\}$ from 402 to 406 nm with $\lambda_{HMM}\approx403$ nm indicated by the dashed line. 
As mentioned earlier,  the primary characteristic of a Type II HMM is increased reflectivity,  shown here as the change in the real part of the ellipsometric parameter $\rho$. 
At the transition wavelength  (dashed line in \ref{fig:RhoSimGraphs}) beyond which the multilayered structure becomes an HMM,  Re$\{\rho\}$ continues to increase with the number of periods until it converges into a band. 
Within the band,  Re$\{\rho\}$ initially increases further, then decreases until stabilizing at a value toward the lower side of the band,  Re$\{\rho_{\infty}\}$ (see simulation details). 
The width of the band is within the confidence interval of $0.035$. 
The fact that the predicted Re$\{\rho\}$ for a \textbf{4P} film sits at the threshold of this band agrees with the observation that the ellipsometric response from the \textbf{4P} and \textbf{7P} films are nearly identical and is interpreted as the emergence of the HMM character,  that is is $P_{HMM,min}= 4$ for a $h_{ML,Ag-Ge}/h_{\ce{DL,Al2O3}}$ combination of 10.84/21.1 nm  ($\Phi=0.34)$ with a 8.82 nm \ce{Al2O3} protection layer.

\subsection{HMM Design Chart}

\begin{figure}[h!]
\begin{subfigure}[]{0.85\textwidth}
        \centering
        \includegraphics[width=\linewidth]{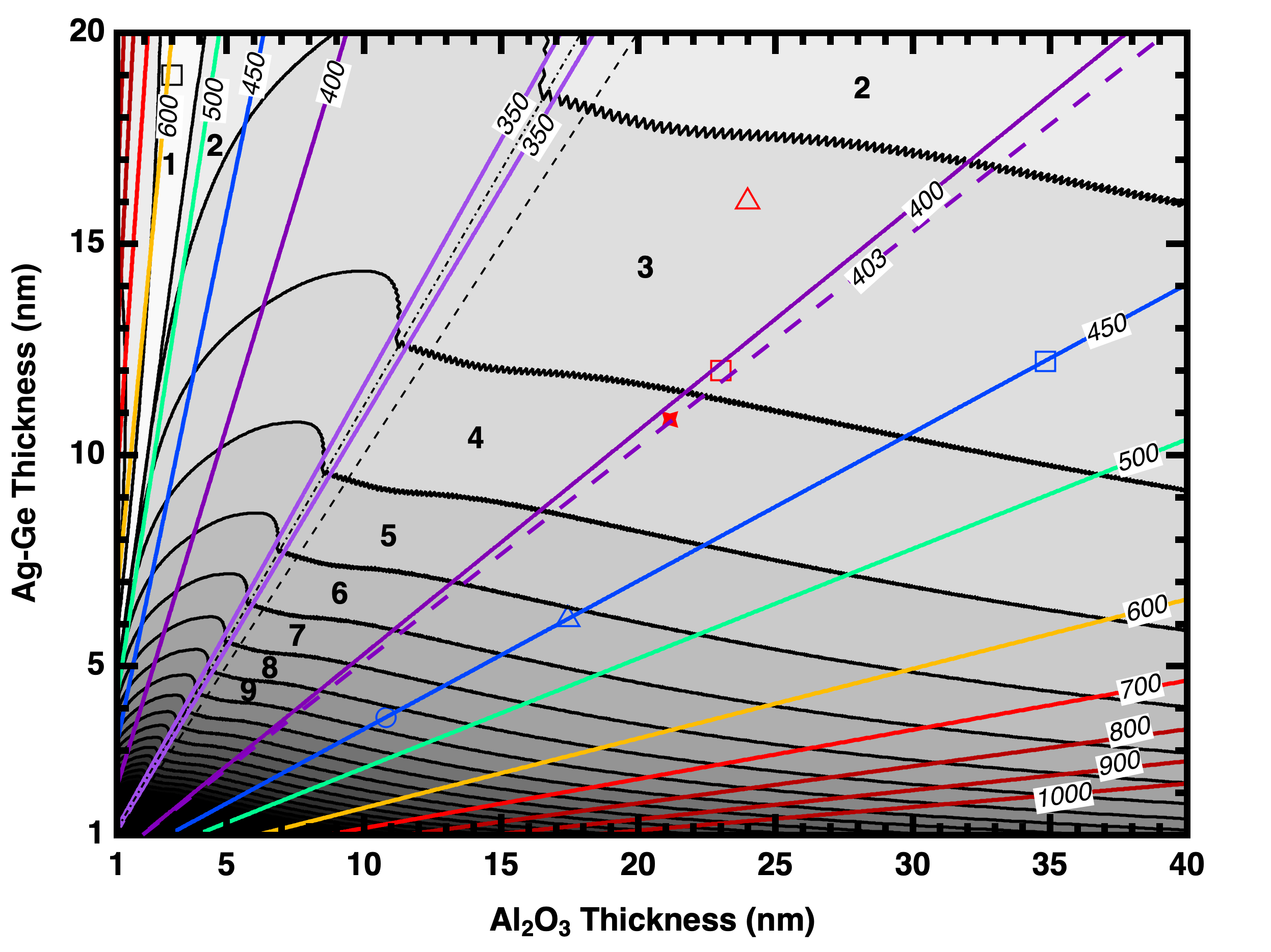}
    
    \end{subfigure}
    \caption{HMM design chart displaying the minimum number of periods,  $P_{HMM.min}$, and the transition wavelength, $\lambda_{HMM}$, needed for a multilayer film comprising alternating layers of $h_{DL,\ce{Al2O3}}$ and $h_{ML,\ce{Ag-Ge}}$ to show Re$\{\rho\}$ within 3.5\% of Re$\{\rho_\infty\}$ based on numerical modeling. 
    Black solid lines indicate the change in $P_{HMM,min}$ by 1 and colored solid lines show selected $\lambda_{HMM}$ values beyond which the film displays hyperbolic dispersion. 
     Black dashed line indicates the 1:1 ratio of \ce{Al2O3}/\ce{Ag-Ge} system, whereas the black dot-dashed line indicates the corrected \ce{Al2O3}/\ce{Ag} ratio (see text).
    Red solid star shows 21.1:10.84 nm \ce{Al2O3}:\ce{Ag-Ge} composition (this work) with  corresponding $\lambda_{HMM}\approx403$ (colored dashed line). 
    Black open square shows a metal-rich point which needs only one period to be an HMM (see text).
    Red open triangle\cite{WS2} and square\cite{Sreekanth} show material compositions discussed in the text from the published literature that show HMM character for three and six periods, respectively. 
    Blue open markers are material combinations shown in \ref{tab:metaldielectriccombos} with $\lambda_{HMM}\approx450$ nm used to explain the utility of the chart (see text).   
    } 
    \label{fig:engineeringchart}
    \hfill
\end{figure}

Using the observed convergence of Re$\{\rho\}$ and the confidence interval ($\delta Re\{\rho\}=0.035$), the numerical analysis is extended into an HMM design chart for other $h_{ML,\ce{Ag-Ge}}/h_{DL,\ce{Al2O3}}$ combinations with a constant \ce{Al2O3} protection layer ($h_{PL,\ce{Al2O3}} = 8.82$ nm) and an incident angle of 65$\degree$ as shown in \ref{fig:engineeringchart}.
The maximum thicknesses for the \ce{Al2O3} and \ce{Ag-Ge} layers in \ref{fig:engineeringchart} are limited to 40 and 20 nm, respectively, as the experimentally obtained thin-film optical constants may not represent the optical constants of the bulk material.\cite{sennettStructureEvaporatedMetal1950, aluminathintobulkPATIL1996120}
\ref{fig:engineeringchart} is obtained by first calculating $\lambda_{HMM}$ for every combination of metal-dielectric thickness using EMT (\response{eq \ref{eq:fillfraction} - 5}). 
Next, the ellipsometry parameter, $\rho$ is simulated for the corresponding $\lambda_{HMM}$.
Last, the period at which $\delta$Re$\{\rho\}$ first falls within 0.035  of Re$\{\rho_{\infty}\}$ (see \ref{fig:RhoSimGraphs}) is identified and denoted as $P_{HMM,min}$. 
Then, $\lambda_{HMM}$ (colored solid lines) and $P_{HMM,min}$ (plateaus in changing gray tones separated by black lines) contour lines are plotted as a function of $h_{DL,\ce{Al2O3}}$  and  $h_{ML,\ce{Ag-Ge}}$. 
Zoomed-in version of \ref{fig:engineeringchart} revealing the details at low silver or dielectric layer thicknesses and additional versions illustrating various aspect of the design chart are provided in SI Section V.

As can be seen in \ref{fig:engineeringchart}, $\lambda_{HMM}$ decreases from the top left corner of the diagram (\ce{Ag-Ge}-rich films), reaches a minimum at $\lambda_{HMM,min} = 348$ nm (dot-dashed black line), and then increases toward the bottom right corner (\ce{Al2O3}-rich films).
$\lambda_{HMM,min}$ coincides with the region in the diagram where the \ce{Ag-Ge} and \ce{Al2O3} layer thicknesses are balanced, i.e., $\Phi\approx 0.5$. 
This symmetry of $\lambda_{HMM}$ as a function of $\Phi$ has been discussed in detail by Cortes et al.\cite{cortesQuantumNanophotonicsUsing2012}
They also show in their work that \ce{Ag}-rich films will transition from an effective metal to a Type II HMM, while \ce{Al2O3}-rich films will make this transition from an effective dielectric. 
In the system discussed here, $\lambda_{HMM,min}$ occurs at $\Phi=0.53$ rather than the expected $\Phi=0.5$ (dashed black line), which is attributed to the presence of the \ce{Ge} seed layer (2 nm nominal thickness) in the system that supports
the growth of smooth silver layers.

$P_{HMM,min}$, which is an integer, increases from the top right corner to the bottom left corner of \ref{fig:engineeringchart} and the solid black lines mark the boundaries at which $P_{HMM,min}$ changes by 1.
A few general trends are observed; (i) as $h_{ML,\ce{Ag-Ge}}$ increases at constant $h_{DL,\ce{Al2O3}}$, $P_{HMM,min}$ decreases (ii) as $h_{DL,\ce{Al2O3}}$ increases at constant $h_{ML,\ce{Ag-Ge}}$, $P_{HMM,min}$ also decreases, and (iii) $P_{HMM,min}$ is more strongly impacted by changes in $h_{ML,\ce{Ag-Ge}}$.
Interestingly, $P_{HMM,min}$ changes monotonically with $h_{ML,\ce{Ag-Ge}}$ for all $h_{DL,\ce{Al2O3}}$, but not so when $h_{DL,\ce{Al2O3}}$ is varied at constant $h_{ML,\ce{Ag-Ge}}$. 
The implication of this feature is that one can make metal-rich and dielectric-rich HHMs with the same $h_{ML,\ce{Ag-Ge}}$ and $P_{HMM.min}$ by varying $h_{DL,\ce{Al2O3}}$ allowing studying the impact of $h_{DL,\ce{Al2O3}}$ on HMM properties.
Another interesting feature, in the metal-rich area of \ref{fig:engineeringchart} is the narrow sliver (very light gray) of material combinations that require only one period ($P_{HMM,min}=1$) to start displaying HMM character,  e.g.,  $h_{ML,\ce{Ag-Ge}}= 19$ nm and $h_{DL,\ce{Al2O3}}=4$ nm (black open square marker, see Figure S13).
A $P_{HMM,min}=1$ is plausible because the presence of the protective \ce{Al2O3} layer creates the second metal/dielectric interface needed for the interaction of two evanescent waves in the film.\cite{elserNonlocalEffectsEffectivemedium2007}

The utility of \ref{fig:engineeringchart} is demonstrated by the red and blue markers.
The red solid star shows the \ce{Ag-Ge} combination used in the \textbf{1P}, \textbf{4P},  and \textbf{7P} films studied here, that is $h_{DL,\ce{Al2O3}}=21.1$ and $h_{ML,\ce{Ag-Ge}}= 10.84$ nm. 
The point falls on the $P_{HMM,min}=4$ plateau and between the $\lambda_{HMM}\approx400$ nm and $\lambda_{HMM}\approx450$ nm lines in good agreement with the similarity in observed ellipsometry response from the \textbf{4P} and \textbf{7P} films (\ref{fig:4P-7P}) and the calculated $\lambda_{HMM}\approx403$ nm (colored dashed line and Figure S7).
The red open triangle shows work by Indukuri et al.,\cite{WS2} in which a three-period \ce{Al2O3}/\ce{Ag} antenna  ($24$/$16$ nm) resulted in a 30-fold enhancement of the overall photoluminescence from a \ce{WS2} monolayer sample agreeing with the $P_{HMM,min}=3$ predicted by \ref{fig:engineeringchart}.
The red open square shows the material combination used for a six-period \ce{Al2O3}/\ce{Ag} film  ($23$/$12$ nm) showing spontaneous emission enhancement for a Coumarin dye,  which is further enhanced by an additional silver grating.\cite{Sreekanth} 
This example also agrees with the HMM design chart in that it has six periods, which is larger than the $P_{HMM,min}=3$ predicted by \ref{fig:engineeringchart}.
Note, both of these works do not use the \ce{Ge} seed layer, which slightly changes the silver optical properties with respect to the reported silver optical constants. 
\ref{fig:engineeringchart} can still be utilized in an \ce{Al2O3}/\ce{Ag} system without the \ce{Ge} seed layer, with a caveat that if the point, falls onto or is close to a $P_{HMM,min}$ plateau boundary, the larger of the two $P_{HMM,min}$ should be used.

Additionally, \ref{fig:engineeringchart} can be utilized to identify practical design constraints, such as the total thickness of the material needed and the time required to fabricate the material.
Using $P_{HMM,min}$, $h_{ML,\ce{Ag-Ge}}$, and $h_{DL,\ce{Al2O3}}$, the minimum total thickness for HMM fabrication denoted as $h_{HMM,min}$ is calculated using \response{eq \ref{eq:hHMM}.}
\begin{equation}
    h_{HMM,min}=P_{HMM,min} (h_{ML,\ce{Ag-Ge}}+h_{DL,\ce{Al2O3}})+h_{PL,\ce{Al2O3}}
    \label{eq:hHMM}
\end{equation}

Three material combinations that exhibit $\lambda_{HMM}\approx 450$ nm are chosen, indicated by the blue markers. 
The material combinations for the three points are summarized in \ref{tab:metaldielectriccombos} together with their total thickness ($h_{HMM,min}$) and the approximated fabrication time  ($\sim t_{fab}$). 
$t_{fab}$ is calculated assuming an average growth rate of $0.3 \textup{~\AA}/s$ and 10 minutes per material change in the evaporator.
Typically,  the choice of HMM fabrication is made based on application,  i.e.,  choosing wavelength and type of HMM. \ref{tab:metaldielectriccombos} shows that the engineering chart can be used to identify several combinations of $h_{DL,\ce{Al2O3}}$ and $h_{ML,\ce{Ag-Ge}}$ that lead to HMMs with $\lambda_{HMM}\approx450$ nm, but have different total film thicknesses, number of periods, and fabrication times.
Therefore,  the combinations chosen for fabrication can now be determined by overall thickness needed for the application or limited by the fabrication time available for production. 
\begin{table}[]
    \centering
    \begin{tabular}{|c|c|c|c|c|c|}
         \hline
         marker&$h_{DL,\ce{Al2O3}}$  (nm)&$h_{ML,\ce{Ag-Ge}}$  (nm)&$P_{HMM,min}$&$h_{HMM,min}$  (nm)&$\sim t_{fab}$  (hours)\\
         \hline
         $\square$&34.84&12.22&3&150&3.06\\
         $\triangle$&17.43&6.1&6&150&4.56\\    
         $\bigcirc$&3.79&10.8&9&140.13&5.96\\ 
        \hline
    \end{tabular}
    \caption{Material thickness combinations for blue open markers in \ref{fig:engineeringchart} with $\lambda_{HMM}\approx450$ nm and corresponding period number ($P_{HMM,min}$), total film thickness  ($h_{HMM,min}$), and approximate time of fabrication  ($\sim t_{fab}$).
    }
    \label{tab:metaldielectriccombos}
\end{table}

\section{Conclusions}
Spectroscopic ellipsometry in conjunction with numerical simulations has been used to identify the onset of HMM character in layered \ce{Ag-Ge/Al2O3} films as a function of number of periods,  $P_{HMM,min}$,  using the convergence of the real part of the ratio of s- and p-polarized reflectivities obtained from \textbf{1P}, \textbf{4P}, and \textbf{7P} films. 
Subsequently,  numerical simulations were used to apply the analysis to HMMs made from \ce{Ag-Ge} and \ce{Al2O3} layers in the range of $1-20$ and $1-40$ nm,  respectively.
The emerging complex HMM design landscape is summarized as a chart that enables identification of number of periods, layer/total thickness, and operational wavelength regime based on application, material, fabrication time, and cost. 
For instance, in super-resolution imaging applications, where ultra-thin layers are usually required to reduce energy losses, one can refer to the bottom left corner of the design chart (see SI Section V, Figure S12).\cite{leeMetamaterialAssistedIllumination2021,leeUltrathinLayeredHyperbolic2022,maSuperresolutionImagingMetamaterialbased2017}
The middle of the design chart is useful in applications controlling spontaneous emission utilizing nano-cavities, where thicker layers are permissible.\cite{maoControllingSpontaneousEmission2024,tumkurControlSpontaneousEmission2011a}
Colloidal metamaterials that rely on core-shell nanoparticles maybe more easily fabricated based on the upper third of the chart that requires low $P_{HMM,min}$.\cite{core-shellexample}
More generally, analogous design charts can be created for other metal-dielectric combinations such as gold/alumina and silver/titania\cite{allHMMsummary}, and is particularly impactful as there is an ongoing search for alternative plasmonic materials that can be more easily synthesized and characterized. \cite{naikAlternativePlasmonicMaterials2013}

\begin{acknowledgement}
The authors acknowledge support from the NSF Phase II CREST Center for Interface Design
and Engineered Assembly of Low Dimensional Materials  (IDEALS),  NSF grant number EES-
2112550. 
IK acknowledges funding through PSC CUNY award 67759-00 55.
This work was performed in part at the Nanofabrication Facility at the Advanced Science Research Center at The Graduate Center of the City University of New York.
\end{acknowledgement}

\begin{suppinfo}
Additional data for ellipsometry calibration and optical constant measurements, details on the numerical simulation, comparison of ellipsometric response from 1P,  4P,  and 7P samples, data for the validation of the numerical simulation, and zoomed-in/additional versions of the HMM design chart.

\end{suppinfo}

\bibliography{HMM_manuscript}

\end{document}